\documentclass[11pt,twoside]{article}

\usepackage{asp2006}
\usepackage{epsf}
\usepackage{lscape}

\begin{document}
\title{Metric radio bursts and fine structures observed on 20 January, 2005}   
\author{C. Bouratzis, P. Preka Papadima, X. Moussas, A. E. Hillaris}   
\affil{University of Athens, Zografos (Athens) , GR-15783, Greece}    
\author{V. Kurt}
\affil{Moscow State University, Moscow, 119992, Russia}  
\author{P.  Tsitsipis, A. Kontogeorgos}
\affil{Technological Educational Institute of Lamia, Lamia , GR-35100, Greece}   

\begin{abstract} 
A  major  radio  event,associated with an X7.1/2B flare in AR720 and a fast CME,
was  observed  on  January  20,  2005  with  the  radio--spectrograph ARTEMIS--IV; 
it was particularly intense and with a complex radio signature with rich fine structure 
which was recorded in the 270--420 MHz range at high 
resolution (100 samples$/$sec). The fine structure is compared with similar results in the 
decimetric and microwave frequency range. It was found to match, almost, the comprehensive Ond{\v r}ejov 
Classification in the spectral range 0.8–-2 GHz.
\end{abstract}
\section*{Data Analysis \& Results}   
\begin{figure}[!ht]
\plotone{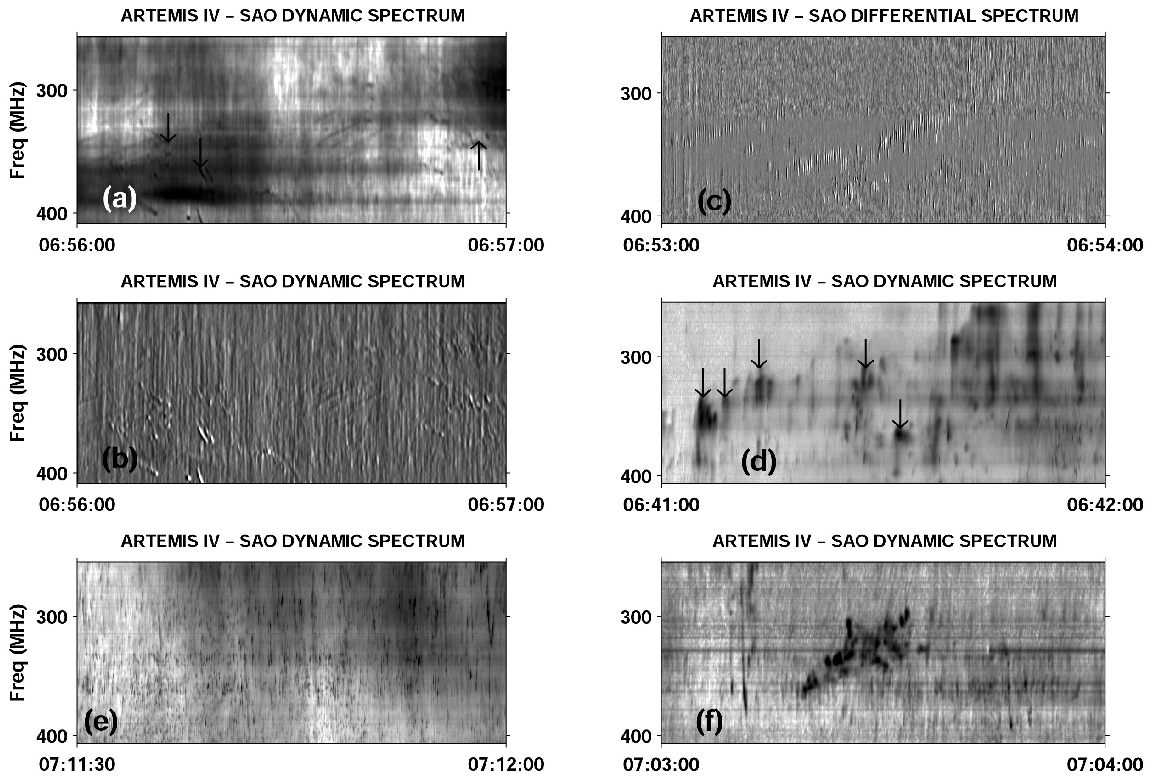}
\plotone{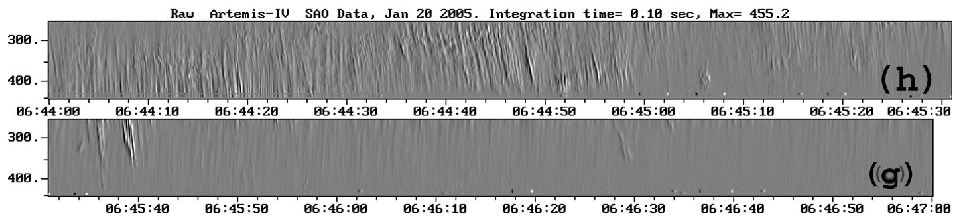}
\caption{(a)\& (b) High Resolution (0.04 sec) Dynamic \& Differential Spectrum of Pulsating structures,
with fibers \& narrowband slow \& fast structures and a Lace Burst (rightmost arrow on (a)). 
(c)Super Short Narrowband Type III (d)Narrowband III(J), III(U) \&  Spikes.
(e) Dot Patches (f)Narrowband III(J)\& III(U) (g)\&(h) Type III(RS) \& Narrowband Type III}
\label{F1}
\end{figure}
In our dynamic spectra, obtained with the by the SAO receiver of the 
ARTEMIS--IV \citep{Caroubalos01,Kontogeorg06,Kontogeorgos08},
the continuum background was removed by means of high-pass filtering on the dynamic spectra 
(differential spectra in this case). The morphological characteristics of fine structure elements embedded in the
metric continuum almost, match the comprehensive Ondrejov Catalogue \citep{Jiricka01} in the  0.8–-2 GHz range.
The high resolution (100 samples$/$s) SAO recordings permitted the recognition and classification of the type III(U) and III(J) subcategory of the narrowband type III bursts in the metric frequency range; similar structures have been reported in the microwaves \citep{Fu04}.

Our observations of fine structures are, briefly, presented in the following subsections; they match, almost, 
similar results in the decimetric and microwave frequency range \citep{Jiricka01,Jiricka02,Fu04,Meszarosova05}.
\subsubsection*{Broadband Pulsations, Fibers \& zebra structures} The pulsations last for hours and form the
background of the dynamic spectra as the type IV continuum has been supressed (cf. figure \ref{F1} (a) \& (b)); 
they are, for most of the same period, associated with fibers. The associated zebra 
structures cover almost the entire pulsation--fiber period. 
\subsubsection*{The narrowband activity} (figure \ref{F1}(a)--(g)) 
included Spikes, Narrow Band Type III (and U) bursts as well as Dot Patches \citep[first reported by][as Super Short Structures]{Magdalenic06}. In figures \ref{F1}(a)\&(b)
a group of slowly \& fast drifting narrow band structures 
outline a \textit{Lace Burst} \citep[cf.][]{Karlicky01}.
 

\end{document}